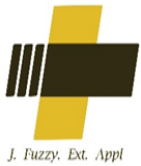

# Journal of Fuzzy Extension and Applications

www.journal-fea.com

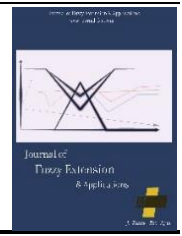

**Paper Type: Research Paper**

# Ranking of Different of Investment Risk in High-Tech Projects Using TOPSIS Method in Fuzzy Environment Based on Linguistic Variables

**Mohammad Ebrahim Sadeghi[1], Hamed Nozari[2,*] , Hadi Khajezadeh Dezfoli[1], Mehdi Khajezadeh Dezfoli[3]**

[1] Department of Industrial Management, Faculty of Management, University of Tehran, Tehran, Iran; sadeqi.m.e@gmail.com; mehkhajezadehd1@gmail.com.

[2] Department of Industrial Engineering, Islamic Azad University, Central Tehran Branch, Tehran, Iran; ham.nozari.eng@iauctb.ac.ir.

[3] Allameh Tabataba'i University, Tehran, Iran; hkhdez@gmail.com.

**Citation:**

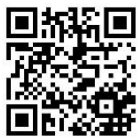

## Abstract

Examining the trend of the global economy shows that global trade is moving towards high-tech products. Given that these products generate very high added value, countries that can produce and export these products will have high growth in the industrial sector. The importance of investing in advanced technologies for economic and social growth and development is so great that it is mentioned as one of the strong levers to achieve development. It should be noted that the policy of developing advanced technologies requires consideration of various performance aspects, risks and future risks in the investment phase. Risk related to high-tech investment projects has a meaning other than financial concepts only. In recent years, researchers have focused on identifying, analyzing, and prioritizing risk. There are two important components in measuring investment risk in high-tech industries, which include identifying the characteristics and criteria for measuring system risk and how to measure them. This study tries to evaluate and rank the investment risks in advanced industries using fuzzy TOPSIS technique based on verbal variables.

## 1 | Introduction

Technology is one of the most important sources of changes in societies. Technology creates new solutions to man problem of daily life [1], [2]. So economic development of countries is intertwined with high tech development [3]. Technology has been a key factor in economic progress over the past few centuries and has played an undeniable role in improving and growing production around the world. New technological developments and constantly changing demands of customers have obliged companies to introduce their new or modified products faster [4].

Technological changes improve the production process of goods and services and increase the efficiency of the production process. In fact, technology change is an integral part of economic growth and development, and in other words, the engine of economic growth; As it is considered the most important factor of economic growth in developed countries in the twentieth century. By emergence of a knowledge-based economy, research productivity, knowledge production, technological innovation, and highly skilled manpower have become key determinants of economic growth [5]. A study of the economic condition of the countries of the world shows that 78% of the economic growth of Germany and 76% of the economic growth of France was due to their technological growth. That figure was 50 percent for the United States. That was 50 percent for the United States. One of the most important features of technology is its increasing return to scale and the fact that its transfer is low cost and after its creation and discovery by one firm, other firms use its knowledge spillover [6].

High-techs considered as the source of technological developments and future industrial revolution, therefore, the economic development and growth of a country in the near future depends on it. That is why many countries have empowered themselves in these technologies and try to utilize its advantages by innovation and technology policy [7]. These technologies are in the early stages of development and it is not possible to make accurate predictions about their development process and dimensions [8]. The importance of investing in high-techs for economic and social development is so great that it is mentioned as one of the strong elements for achieving development [9]. But it should be noted that just as paying attention to this can cause economic growth by shaping a virtues cycle, not paying attention to it may lead to economic decline and falling into a vicious cycle. Therefore, economic growth and increasing public welfare in the long term will not be possible without investing in these industries and paying attention to the risks associated with it.

Investment has two components of risk and return, and the relationship between them offers different combinations of investment. On the one hand, investors seek to maximize their return on investment, and on the other hand, they face conditions of uncertainty in the market and industry environment, which makes investing uncertain, therefore, success is not guaranteed. Of every 7 to 10 innovative product concepts, only one will gain commercial success. Also, forty percent of innovative products fail at launch, despite their successful development and passing performance tests [10]. In this regard many researchers expressed that uncertainty affects not only on real economic activity but also on the investment decisions of economic agents [11]. In other words, all investment decisions are based on the relationship between risk and return. However, the high volume of global trade in high-tech products motivates investors to enter these industries. However, it should be noted that the policy of high techs requires consideration of various dimensions and aspects of performance and risks in the investment process. These products require advanced technologies that are changing rapidly. It also requires adequate infrastructure, highly skilled human resources, and strong links between firms and relation between firms and research centers and universities.

In order to invest in these industries, two important factors must be paid special attention: firstly, these industries need high investments and secondly, the investment processes in these industries face complex risks. Risks in high-tech investment projects have dimensions other than financial ones only. There are risks in technological, competitive, managerial aspects and some other risk arising from the presence of asymmetric information [12]. In the process of investing and implementing high-tech projects, events may occur that jeopardize the occurrence, implementation and profitability of the project. Therefore, identifying, analyzing, prioritizing and having a plan to deal with these events, can play an important role in the success of the investment project in high techs. In recent years, researchers have focused on identifying, analyzing, and prioritizing risk in high tech. There are two important components in measuring investment risks in high-tech industries, which include determining the criteria for measuring system risk and how to measure them. In this research, we tried to evaluate and rank the investment risks in high tech industries with fuzzy TOPSIS technique based on linguistic variables.

## 2 | Literature Review

Studies on the risk of investing in high-tech projects date back to the 1960s. Numerous experimental studies have been performed by Myers and Marquitz. Their studies focused more on financial metrics, but market uncertainty and project technology were not considered [13]. Until in 1970, the criteria of technology, market and management were examined in related research. researchers divided the evaluation criteria into four categories: production, company capacity, environmental factors, and alternative competition. Other scholars Stated that venture capitalists should consider the five areas of skill, technology, production, market, and investment to evaluate a new investment project [14]. Based on the qualitative evaluation criteria, Tyebjee and Bruno [15], for the first time, used a questionnaire method to identify evaluation factors to structure the investment risk assessment model in US projects. They selected the 12 evaluation criteria that investors often referred to in their questionnaires. Fried and Hisrich [16] set 15 initial evaluation criteria and divided them into three areas: strategic thinking, management capacity, and research revenue. Manigart et al. [17] first interviewed and researched a number of venture capital project investors from the United Kingdom, Ireland, Belgium and France. They chose factors that affect investment income and investment risk. The team found that the risk of the firm and the target market management team had the greatest impact on investment risk, production innovation, and expected return on investment. They believed that the general economic situation had the least effect on the rate of return [17]. Chotigeat et al. [18] studied the risk assessment criteria for investment in Sri Lanka and Thailand and found that those countries have their own investment assessment criteria. Sri Lankan empirical studies have shown that venture capitalists first emphasize the future return on investment of the firm and then examine market demand, management team, market growth potential and investment liquidity. Venture capitalists in Thailand, on the other hand, first looked at the capacity of the management team and then at the return on investment. Kaplan and Stromberg's study highlighted the criteria for investment attractiveness (market size, strategy, technical, customer), competitiveness, and the subject matter of investment, which is often of interest to investors [19]. The study of the investment risk of Zutshi et al. [20] in Sri Lanka shows that entrepreneurial personality is the most important factor in evaluating investment and financial factors are the least important. To create an investment analysis system, researchers examined investment risks, which resulted in the formation of a 12-factor valuation system that falls into five areas: product differentiation, market attractiveness, capacity, and management capacity. Economic effectiveness, and environmental impact. Given the real situation in China where investing companies could not withdraw their capital, the investment risk decision model proposed by previous researchers modified and a new factor called the exit strategy, which included two exit strategies added to it. These two factors included the degree of difficulty or ease of withdrawing capital and the method of withdrawing capital. Some scholars proposed a combined assessment system that includes three subsystems: environmental assessment, risk assessment, and economic income assessment, and proposed 50 criteria and factors. Han [21] proposed six indicators of technical risk, production risk, market risk, operational risk, financial risk, and environmental risk, which include 26 two-tier criteria. Qiu-bai [22] divide the investment risks in the project into systematic and non-systematic risks. Systematic risks include political and social risks, economic risks and legal risks. Unsystematic risks include technical risk, production risk, management risk, financial risk and market risk.

Um and Kim [23] express task uncertainty as one of the most important uncertainties in innovative projects. They identified three major causes of innovation project task uncertainty which include: product complexity, technological novelty, and task interdependence. Fanousse et al. [24] by integrating the previous studies, indentify 11 main uncertainties in innovation projects. Three of them are the main uncertainties, emphasized by most scholars that are technological, market and task. Other 8 identified uncertainties are perceived environmental, regulatory/institutional, perceived social, collaboration, organisational, decision, financial, operational and ontological. Also, it is obvious that there are many differences between the environments and systems of different regions in different countries. Therefore, investment evaluation criteria are not the same in different countries [13].

Scholars used fuzzy mathematical methods and the Analytic Hierarchy Process to create a quantitative model for evaluating investment risks in high-tech projects [25]. Scholars used fuzzy mathematical methods and the Analytic Hierarchy Process to create a quantitative model for evaluating investment risks in high-tech projects [13]. Tang et al. [26] by considering mutual influence between factors use fuzzy network analysis to evaluate risks in urban rail transit projects. Zhao and Li [27] in their study of key risk factors of Ultra-High Voltage projects suggest a risk index structure. Their research is based on a cloud model and Fuzzy Comprehensive Evaluation (FCE) method. They combine the superiority of the cloud model for reflecting randomness and discreteness with the advantages of the fuzzy comprehensive evaluation method in handling uncertain and vague issues. Wang et al. [28] in their study of risk ranking of energy performance contracting project, develop a multi-criteria decision-making framework under the picture fuzzy environment model. In order to create a risk assessment framework, Wu et al. [29] use linguistic hesitant fuzzy sets based cloud model, for seawater pumped hydro storage project under three typical public-private partnership management modes. Baylan [30] combined AHP and TOPSIS methods to develop a multi-criteria based decision method that prioritizes project risks at the activity level.

In Iran, in the last two decades, many high techs projects have been launched. Both public and private sectors made a lot of investments in advanced technologies. However, many studies investigate risks in projects (for example see [31]), few studies appear to identify and rank investment risks in high-tech industries (for example see [32]). Of course, many positive moves have been made in the country to support innovative projects, such as the establishment of technology parks ([33], [34]) and venture capital funds, but there is still a gap in studies related to the identification and ranking of investment risks in the field of high technologies.

## 3 | Theoretical Framework of Research Method

There are various methods for ranking factors in different studies. The most famous method is the family of Multi-Criteria Decision Making (MCDM). The MCDM methods are among the best methods in dealing with decision-making problems [35]. Multi-criteria decision-making techniques are divided into two categories: MODM like TOPSIS, AHP [36] and MADM) like SAW [37]. MCDM techniques and group decision making have a wide range of applications in the literature and allow managers and decision makers to evaluate options in different dimensions. MCDM includes various techniques such as TOPSIS, AHP, etc. These methods are widely used due to their practicality, so that today their use has spread to all fields and disciplines [38].

The TOPSIS technique is one of the most popular classic MCDM techniques first introduced by Hwang and Yoon [39]. The basic logic of TOPSIS is the definition of ideal and anti-ideal solution. The ideal solution is a solution that maximizes positive criteria and minimizes negative criteria. the ideal solution contains all the best values of the available criteria, while the anti-ideal solution is a combination of the worst values of the available criteria. The optimal option is the option that has the shortest distance from the ideal solution and the longest distance from the anti-ideal solution [40]. Because TOPSIS is a popular method for classical MCDM problems, many researchers use it to solve ranking and prioritization problems. In fact, TOPSIS is a practical method that compares alternatives according to their values in each criterion and the weight of the criterion [41].

However, in many cases it is not possible to measure values with any particular degree of accuracy. Hence, inaccuracies occur in the information obtained. The sources of imprecision are unquantifiable information, incomplete information, non-obtainable information, and partial ignorance [42]. It should be noted that the problem is not a lack of information, but there is uncertainty in the information. Such uncertainty can be formulated with non-random intervals. In fact, these uncertainties can be easily modeled with fuzzy sets. [43]

The fuzzy logic method was first proposed by Zadeh [44]. There are many inaccurate concepts around us that are expressed on a daily basis in the form of various phrases. Fuzzy logic is a new process that replaces the methods that require advanced and sophisticated mathematics to design and model a system with linguistic quantities and expert knowledge [44]. Zadeh argues that humans do not have a lot of accurate information inputs, but are able to perform adaptive control extensively [45]. In fact, fuzzy logic provides an easy way to reach a definite result based on incomplete, erroneous, ambiguous, vague input information. In this regard, in the present study, TOPSIS technique in fuzzy environment is used to prioritize the factors affecting the investment risk in high-tech industries.

In the classical TOPSIS method, accurate and definite numerical values are used to rank the alternatives and determine the weight of each criterion. But it is not always possible for decision makers to express their thoughts and decisions accurately and quantitatively, so they use linguistic variables such as good, bad, poor, etc. to reflect their opinions. In such cases, it is possible to use the theory of fuzzy sets to express the views and evaluate the opinions of decision makers. The fuzzy TOPSIS algorithm is one of the efficient algorithms in the category of multi-criteria decision making problems in which the elements of the decision matrix or the weight of the criteria or both are expressed by linguistic variables. The important point in the ranking process is that the metrics of this model are expressed in terms of subjective, qualitative and linguistic variables [46].

Fuzzy set theory and fuzzy logic as mathematical theories are a very efficient and useful tool for modeling and formulating mathematical ambiguity and inaccuracy in human cognitive processes. Fuzzy set theory provides tools that mathematical formulate human reasoning and decision making mathematically, so these mathematical models can be used in a variety of fields of science and technology [47]. Using fuzzy concepts, evaluators can use verbal expressions in colloquial natural language to evaluate effective factors, and by linking these expressions to appropriate membership functions, analysis of scores and components will be more appropriate and accurate [48].

## 3.1 | Overview of TOPSIS Method and Fuzzy Calculations

In a multi-criteria decision making problem with m options and n criteria and using triangular fuzzy numbers, the following steps are used to rank the options[35], [47], [49]- [51]:

**Step 1. creating decision matrix.**

In the first step, we create the decision matrix according to the criteria and options:

$$\widetilde{D} = \begin{bmatrix} \tilde{x}_{11} & \tilde{x}_{12} & \cdots & \tilde{x}_{1n} \\ \tilde{x}_{21} & \tilde{x}_{22} & \cdots & \tilde{x}_{2n} \\ \vdots & \vdots & \ddots & \vdots \\ \tilde{x}_{m1} & \tilde{x}_{m2} & \cdots & \tilde{x}_{mn} \end{bmatrix} \qquad (1)$$

Where $\tilde{x}_{ij}$ is a triangular fuzzy number corresponding to the ith option according to the $j_{th}$ criterion and $i = 1,2,\ldots, m$ and $j = 1,2,\ldots, n$.

If there are K decision makers and the fuzzy ranking of the kth decision maker for $i = 1,2,\ldots, m$ and $j = 1,2,\ldots, n$ as a triangular fuzzy number is $\tilde{x}_{ijk} = a_{ijk}, b_{ijk}, c_{ijk}$ then the combined fuzzy ranking of decision makers' opinions about options is $\tilde{x}_{ij} = a_{ij}, b_{ij}, c_{ij}$ and can be obtained based on the following relationships:

$$a_{ij} = \text{Min}_K\{a_{ijk}\},$$
$$b_{ij} = \frac{\sum_{k=1}^{K} b_{ijk}}{K}, \qquad (2)$$
$$c_{ij} = \text{Max}_K\{c_{ijk}\}.$$

Table 1. Relation of triangular numbers with linguistic variables.

| Linguistic Variables | Fuzzy Triangular Numbers According to Linguistic Variables |
|---|---|
| Very low | (1, 1, 3) |
| weak | (1, 3, 5) |
| medium | (3, 5, 7) |
| high | (5, 7, 9) |
| Very high | (7, 9, 9) |

**Step 2. Determining the weighted criteria matrix.**

If we consider the fuzzy matrix $\widetilde{W} = [\widetilde{w}_1, \widetilde{w}_2, \ldots, \widetilde{w}_n]$ as the weighted criteria matrix and $\widetilde{w}_j$ be a triangular fuzzy number as $\widetilde{w}_j = w_{j1}, w_{j2}, w_{j3}$ .and If the minimum number of decision makers is equal to K and the Kth decision maker's significance coefficient be a triangular fuzzy number as $\widetilde{w}_{jk} = w_{jk1}, w_{jk2}, w_{jk3}$ for $j = 1,2,\ldots, n$, then the combined fuzzy ranking $\widetilde{w}_j = w_{j1}, w_{j2}, w_{j3}$ can be obtained by using the following equations:

$$w_{j1} = \text{Min}_K\{w_{jk1}\},$$

$$w_{j2} = \frac{\sum_{k=1}^{K} w_{jk2}}{K}, \tag{3}$$

$$w_{j3} = \text{Max}_K\{w_{jk3}\}.$$

**Step 3. make the decision matrix dimensionless.**

In this step, linear scale transformation is used to make the fuzzy decision matrix dimensionless so the comparison of different options is comparable. The components of the dimensionless decision matrix for positive and negative criteria are calculated from the following equations, respectively:

$$\widetilde{r}_{ij} = \left(\frac{a_{ij}}{c_j^*}, \frac{b_{ij}}{c_j^*}, \frac{c_{ij}}{c_j^*}\right),$$

$$\widetilde{r}_{ij} = \left(\frac{a_j^-}{c_{ij}}, \frac{a_j^-}{b_{ij}}, \frac{a_j^-}{a_{ij}}\right), \tag{4}$$

$$c_j^* = \text{Max}_i c_{ij},$$

$$a_j^- = \text{Min}_i a_{ij}.$$

According to the above steps, the dimensionless fuzzy decision matrix is obtained as follows:

$$\widetilde{R} = [\widetilde{r}_{ij}]_{m \times n}, i=1,2,\ldots,m ; j=1,2,\ldots,n$$

$$\widetilde{R} = \begin{bmatrix} \widetilde{r}_{11} & \widetilde{r}_{12} & \cdots & \widetilde{r}_{1n} \\ \widetilde{r}_{21} & \widetilde{r}_{22} & \cdots & \widetilde{r}_{2n} \\ \vdots & \vdots & \ddots & \vdots \\ \widetilde{r}_{m1} & \widetilde{r}_{m2} & \cdots & \widetilde{r}_{mn} \end{bmatrix}. \tag{5}$$

Where m and n represent the number of options and the number of criteria, respectively.

**Step 4. Obtain the weighted decision matrix.**

The weighted decision matrix is calculated by multiplying the coefficient of significance related to each of the criteria in the fuzzy scaleless matrix and the calculation method for the positive and negative criteria is as follows:

$$\tilde{v}_{ij} = \tilde{r}_{ij} \cdot \tilde{w}_j = \left(\frac{a_{ij}}{c_j^*}, \frac{b_{ij}}{c_j^*}, \frac{c_{ij}}{c_j^*}\right) \cdot (w_{j1}, w_{j2}, w_{j3}) = \left(\frac{a_{ij}}{c_j^*} \cdot w_{j1}, \frac{b_{ij}}{c_j^*} \cdot w_{j2}, \frac{c_{ij}}{c_j^*} \cdot w_{j3}\right).$$

$$\tilde{v}_{ij} = \tilde{r}_{ij} \cdot \tilde{w}_j = \left(\frac{a_j^-}{c_{ij}}, \frac{a_j^-}{b_{ij}}, \frac{a_j^-}{a_{ij}}\right) \cdot (w_{j1}, w_{j2}, w_{j3}) = \left(\frac{a_j^-}{c_{ij}} \cdot w_{j1}, \frac{a_j^-}{b_{ij}} \cdot w_{j2}, \frac{a_j^-}{a_{ij}} \cdot w_{j3}\right).$$

(6)

Where $\tilde{w}_j$ is the significance factor of the criterion $j$.

According to the above, we will have:

$$\tilde{V} = [\tilde{v}_{ij}]_{m \times n}, i=1,2,\ldots,m \; ; j=1,2,\ldots,n$$

$$\tilde{V} = \begin{bmatrix} \tilde{v}_{11} & \tilde{v}_{12} & \cdots & \tilde{v}_{1n} \\ \tilde{v}_{21} & \tilde{v}_{22} & \cdots & \tilde{v}_{2n} \\ \vdots & \vdots & \ddots & \vdots \\ \tilde{v}_{m1} & \tilde{v}_{m2} & \cdots & \tilde{v}_{mn} \end{bmatrix}.$$

(7)

**Step 5. Determine the positive and negative ideal solution.**

Positive ideal and negative ideal solutions are defined as follows:

$$A^* = \{\tilde{v}_1^*, \tilde{v}_2^*, \ldots, \tilde{v}_n^*\},$$

$$A^- = \{\tilde{v}_1^-, \tilde{v}_2^-, \ldots, \tilde{v}_n^-\}.$$

(8)

Where $\tilde{v}_i^*$ is the best value of criterion i among the options and $\tilde{v}_i^-$ is the worst value of criterion $i$ among all available options. In fact, in this step we want to find the best and worst possible option.

**Step 6. Distance from the ideal positive and negative fuzzy solution.**

These distances can be calculated according to the following equations:

$$S_i^* = \sum_{j=1}^n d(\tilde{v}_{ij}, \tilde{v}_j^*) \; , \quad i=1,2,\ldots,m$$

$$S_i^- = \sum_{j=1}^n d(\tilde{v}_{ij}, \tilde{v}_j^-) \; , \quad i=1,2,\ldots,m$$

(9)

Thus, the distance between two triangular fuzzy numbers $(a_1, b_1, c_1)$ and $(a_2, b_2, c_2)$ is calculated as follows:

$$d_v(\tilde{M}_1, \tilde{M}_2) = \sqrt{\frac{1}{3}[(a_1 - a_2)^2 + (b_1 - b_2)^2 + (c_1 - c_2)^2]}.$$

(10)

**Step 7. Calculate the similarity index.**

Similarity index can be calculated according to the following equation:

$$CC_i = \frac{S_i^-}{S_i^- + S_i^*}$$

(11)

**Step 8. Ranking the options.**

At this point the options are ranked according to the $CC_i$ values. So, the options that have a higher similarity index will have better rankings.

# 4 | Methods and Results of Research

The present study seeks to identify and prioritize significant risks in investing in high-tech industries. Therefore, in line with the objectives of the research, first by reviewing the literature and measures done in other countries, as well as using the opinions of experts, thirty factors affecting investment risk in high-tech industries are identified and classified into six general categories including financial risk and technology risk, production risk, market risk, management risk, and environment risk.

Finally, these factors and components through a questionnaire and obtaining the opinions of experts (30 people) were measured by verbal variables. The experts have a master or PhD degree in financial management, metallurgical engineering, industrial engineering, chemical engineering (nanotechnology) and financial engineering from Tehran University, Iran University of Science and Technology, Amirkabir University, Tarbiat Modarres University, Allameh Tabatabai University and Economic Sciences University.

Also, some Experts of the capital and financial market activists of the country were included to the survey. It is necessary to explain that the selection of the number of people from each specialty as well as the type of specialization is based on the nature of high-tech projects and the type of relevant questions.

**Table 2. Identified risks in investing in high-tech projects.**

| Type of Risk | |
|---|---|
| financial risk | Financial capability |
| | Ability to raise production capital |
| | Change in interest rates |
| | Change the exchange rate |
| | Capital market volume |
| Technology risk | Technological advantage |
| | Technological maturity |
| | Reliability of technology |
| | Alternative technology |
| | Professional work experience |
| Production risk | How difficult or easy it is to work with technology |
| | how standard the equipment and production process are |
| | Employee decisions |
| | Raw material supply capacity |
| | Raw material prices |
| | Product life cycle |
| Market risk | Capacity and time of admission |
| | Product competitiveness |
| | Potential rival effect |
| | Marketing capability |
| | Network readiness |
| | New technology acceptance network |
| Management risk | Quality and experience of managers |
| | The ease of obtaining information |
| | The rate of use of collective wisdom |
| | Project management mechanism |
| Environment risk | The desirability of legal environment policies |
| | Macroeconomic environment desirability |
| | Favorable social environment |
| | the environment condition |

In the next step, the information obtained from the questionnaires was extracted and after entering them in EXCEL software and performing the fuzzy TOPSIS algorithm, the following results were extracted:

Table 3. The values of the ideals.

| Positive Ideal | (1,1,1) |
|---|---|
| Negative Ideal | (0/111,0/111,0/111) |

Table 4. Results extracted from questionnaires.

| Types of Risks | Fuzzy Average of Expert Opinions | The Distance from The Positive Ideal | The Distance from The Negative Ideal | Similarity Index | Ranking |
|---|---|---|---|---|---|
| Financial capability | (3,7/133,9) | 0.403 | 0.659 | 0.62 | 1 |
| Ability to raise production capital | (3,6/866,9) | 0.408 | 0.648 | 0.613 | 2 |
| Change in interest rates | (1,6/266,9) | 0.542 | 0.614 | 0.531 | 8 |
| Change the exchange rate | (1,5/6,9) | 0.557 | 0.592 | 0.515 | 11 |
| Capital market volume | (1,4/466,9) | 0.59 | 0.559 | 0.486 | 21 |
| Technological advantage | (1,5/466,9) | 0.561 | 0.587 | 0.511 | 12 |
| Technological maturity | (1,5/066,9) | 0.572 | 0.575 | 0.501 | 16 |
| Reliability of technology | (1,5/066,9) | 0.572 | 0.575 | 0.501 | 16 |
| Alternative technology | (3,6/866,9) | 0.408 | 0.648 | 0.613 | 2 |
| Professional work experience | (1,6/4,9) | 0.539 | 0.619 | 0.534 | 6 |
| How difficult or easy it is to work with technology | (1,5/066,9) | 0.572 | 0.575 | 0.501 | 17 |
| how standard the equipment and production process are | (1,5/4,9) | 0.562 | 0.585 | 0.51 | 13 |
| Employee decisions | (1,4/666,9) | 0.583 | 0.564 | 0.491 | 19 |
| Raw material supply capacity | (1,5/133,9) | 0.57 | 0.577 | 0.503 | 16 |
| Raw material prices | (1,5/066,9) | 0.572 | 0.575 | 0.501 | 16 |
| Product life cycle | (1,6/2,9) | 0.543 | 0.611 | 0.529 | 9 |
| Capacity and time of admission | (1,5/866,9) | 0.551 | 0.6 | 0.521 | 10 |
| Product competitiveness | (1,6/533,9) | 0.537 | 0.623 | 0.537 | 5 |

**Table 4. (Continued).**

| Types of Risks | Fuzzy Average of Expert Opinions | The Distance from The Positive Ideal | The Distance from The Negative Ideal | Similarity Index | Ranking |
|---|---|---|---|---|---|
| Potential rival effect | (1,6/266,9) | 0.542 | 0.614 | 0.531 | 8 |
| Marketing capability | (1,6/6,9) | 0.535 | 0.626 | 0.539 | 4 |
| Network readiness | (1,5/266,9) | 0.566 | 0.581 | 0.506 | 14 |
| New technology acceptance network | (1,5,9) | 0.573 | 0.573 | 0.5 | 18 |
| Quality and experience of managers | (1,6/333,9) | 0.541 | 0.616 | 0.532 | 7 |
| The ease of obtaining information | (1,5/2,9) | 0.568 | 0.579 | 0.505 | 15 |
| The rate of use of collective wisdom | (1,5,9) | 0.573 | 0.575 | 0.501 | 17 |
| Project management mechanism | (1,5/2,9) | 0.568 | 0.579 | 0.504 | 15 |
| The desirability of legal environment policies | (1,5/2,9) | 0.568 | 0.579 | 0.504 | 15 |
| Macroeconomic environment desirability | (1,6/733,9) | 0.533 | 0.631 | 0.542 | 3 |
| Favorable social environment | (1,4/533,9) | 0.588 | 0.56 | 0.487 | 20 |
| the environment condition | (1,3/333,9) | 0.629 | 0.534 | 0.459 | 22 |

In the final stage, using the fuzzy average method and the above information, six risk categories were ranked and the following results were obtained:

**Table 5. Ranking of six risks.**

| Types of Risks Studied | Fuzzy Average of Expert Opinions | The Distance from the Positive Ideal | The Distance from the Negative Ideal | Similarity Index | Ranking |
|---|---|---|---|---|---|
| financial risk | (1,6/066,9) | 0/546 | 0/607 | 0/526 | 1 |
| Technology risk | (1,5/773,9) | 0/553 | 0/597 | 0/519 | 3 |
| Production risk | (1,5/066,9) | 0/571 | 0/575 | 0/502 | 5 |
| Market risk | (1,5/85,9) | 0/551 | 0/6 | 0/521 | 2 |
| Management risk | (1,5/433,9) | 0/562 | 0/586 | 0/510 | 4 |
| Environment risk | (1,4/95,9) | 0/575 | 0/572 | 0/499 | 6 |

## 5 | Conclusion

Today, the trend of the global economy reflects the trend of trade towards products with advanced technologies. Naturally, countries that can produce and export these products will have high growth in the industrial sector, and on the contrary, neglecting it can cause economic decline in the future. Given the growing importance of these products in world trade, it is necessary to move to expand investment in these technologies in the country. The findings of this study show that investing in high-tech industries faces many risks. If we divide these risks into six categories: financial risk, market risk, environmental risk, technology risk, production risk and management risk, based on the opinion of elites and experts and using the scientific method of TOPSIS evaluation model in fuzzy environment based on verbal variables. The degree of importance of these risks in the country is in the form of financial risk, market risk, technology risk, management risk, production risk and finally environmental risk. This ranking and information reflect the opinion of the elites about the state of the country to invest in high-tech projects. Naturally, paying attention to it is necessary for principled planning and accurate policy-making in the country, as well as the awareness of domestic and foreign investors about the investment climate in the country.

## Conflicts of Interest

All co-authors have seen and agree with the contents of the manuscript and there is no financial interest to report. We certify that the submission is original work and is not under review at any other publication.